# Acoustic Wave Induced FMR Assisted Spin-Torque Switching of Perpendicular MTJs with Anisotropy Variation

Walid Al Misba[1], Md. Mahadi Rajib[1], Dhritiman Bhattacharya[1], Jayasimha Atulasimha[1,2*]

[1]Dept. of Mech. & Nuc. Engineering, Virginia Commonwealth University, Richmond, VA, US

[2]Dept. of Elec. & Comp. Engineering, Virginia Commonwealth University, Richmond, VA, US

**Abstract:**

We have investigated Surface Acoustic Wave (SAW) induced ferromagnetic resonance (FMR) assisted Spin Transfer Torque (STT) switching of perpendicular MTJ (p-MTJ) with inhomogeneities using micromagnetic simulations that include the effect of thermal noise. With suitable frequency excitation, the SAW can induce ferromagnetic resonance in magnetostrictive materials, and the magnetization can precesses in a cone with high deflection from the perpendicular direction. With incorporation of inhomogeneity via lateral anisotropy variation as well as room temperature thermal noise, the magnetization precession in different gains can be significantly incoherent. Interestingly, the precession in different grains are found to be in phase, even though the precession amplitude (angle of deflection from the perpendicular direction) vary across grains of different anisotropy. Nevertheless, the high mean deflection angle can complement the STT switching by reducing the STT current significantly; even though the applied stress induced change in anisotropy is much lower than the total anisotropy barrier. This work indicates that SAW assisted switching can improve energy efficiency while being scalable to very small dimensions, which is technologically important for STT-RAM and elucidates the physical mechanism for the potential robustness of this paradigm in realistic scenarios with thermal noise and material inhomogeneity.

**Introduction:**

Magnetic tunnel junctions (MTJ) are finding increasing application as non-volatile nanomagnetic memory devices, which are an alternative to volatile CMOS based memory devices. The most prevalent scheme to accomplish magnetization switching of the free layer of an MTJ (i.e. writing bits) utilizes spin transfer torque (STT) [1, 2]. Although STT memory can be scaled to ~10 nm, the energy requirement has not decreased below 100 fJ/bit [3]. Therefore, alternative strategies such as strain mediated [4, 5] and voltage mediated MTJ [6, 7] switching have been investigated. However, scaling of strain-based devices is challenging. As the volume (V) shrinks, to maintain an energy barrier with sufficient thermal stability ($E_b = K_u V \sim 1eV$), large perpendicular anisotropy ($K_u$) is required. In such a scenario, the static stress ($\sigma$) required to erode the energy barrier, which is determined by $E_{stress} = 3/2\, \lambda_s \sigma V = E_b$, would also be very large. For example, to erode an energy barrier of $E_b$ ~1eV in a circular nanostructure with lateral dimension of 20 nm and thickness 1 nm (i.e. volume, V ~314 nm$^3$), and a saturation magnetostriction of $\lambda_s$ = 200 ppm the stress amplitude could be as high as 1.7 GPa. This is possibly an order of magnitude higher than the stress that can be generated dynamically or applied for many cycles in a practical device.

In contrast to static stress, time varying stress can drive the magnetization to acoustically induced ferromagnetic resonance (A-FMR) [8]. This can lead to large amplitude magnetization precession even when the stress induced anisotropy change is substantially smaller than the total energy barrier as the amplitude of this precession grows due to the energy added over many cycles. Physically, time varying strain can be generated by surface acoustic waves (SAW) via interdigital SAW electrode deposited and patterned over a piezoelectric substrate. Previously, SAW driven FMR on Ni film [9] and magnetization switching for magnetostrictive Co nanomagnets has been experimentally reported [10]. Precessional magnetization switching [11] with SAW on (GA, Mn) (AS, P) film and field free switching with SAW for (Ga, As) P [12] has also been experimentally reported at low temperature. Laser pump induced SAW and their magnetization dynamics has been studied for single nanomagnet [13] and on patterned periodic nanodots [14] and their magnetization reversal has been numerically investigated in nanomagnets [15].

SAW assisted STT induced magnetization reversal for in-plane and perpendicular MTJ based on simulations using macro-spin assumption has recently been reported [16]. The main purpose of the scheme is to induce magnetization rotation with SAW so that when the STT is applied, the magnetization experiences higher torque. However, macrospin assumption precludes modeling of incoherence in the magnetization dynamics, which could arise due to inhomogeneity in material properties. This inhomogeneity can be an intrinsic material property [17], or due to edge modifications [18], roughness variation [19], thickness variation over an extended area [20], etc. While the incoherence in magnetization mainly stems from the competition between the long ranged weak magnetostatic energy and short ranged strong exchange energy and the balance is ultimately decided by the size and shape of the nanomagnetic structure; the above-mentioned inhomogeneities can increase the incoherency even in smaller size nanomagnet by varying the local anisotropy field.

In this study, we perform micromagnetic simulations that incorporate the incoherent magnetization dynamics in the presence of room temperature thermal noise as well as lateral variation in the uniaxial (perpendicular) anisotropy. Furthermore, the cell size of lateral dimensions 1.56 nm × 1.56 nm naturally includes an edge roughness ~ 1 nm that is consistent with lithography and fabrication limitations. It is expected that all these realistic variations could lead to incoherent magnetization precession. For example, the magnetization in different grains (Voronoi tessellation was used to create regions with an average lateral dimension ~ 10 nm) can precess at different cones when driven by a SAW as shown in Fig. 1c. This incoherent (non-uniform) magnetization precession reduces the net magnetization of the nanomagnet (as in Fig. 1c although some regions precess in high cone, the net magnetization remains very low) and in extreme cases reduces it to zero. Therefore, we study this incoherent precession, particularly in the SAW induced FMR regime, to understand the underlying magnetization dynamics as well as its ramifications on the ability to significantly reduce the STT switching current. This has important implications towards implementation of energy efficient STTRAM that are scalable to small lateral dimensions.

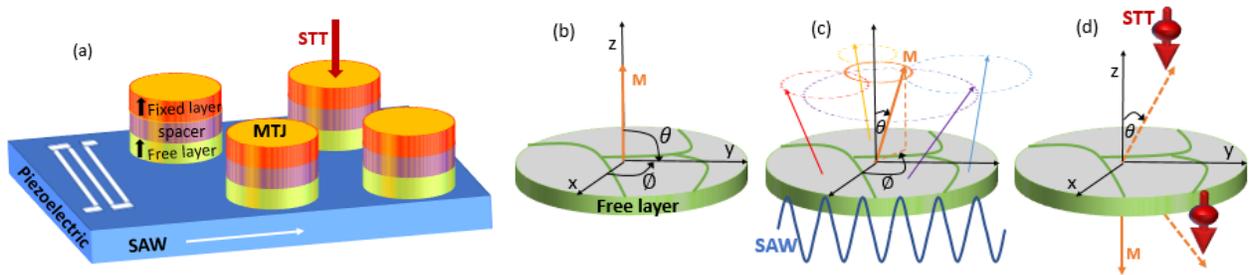

FIG. 1. a. MTJ arrays and SAW electrode over piezoelectric substrate b. initial magnetization state of the inhomogeneous (i.e. granular) free layer c. application of SAW induces different angle precession and the resulting incoherency reduces the net magnetization, **M** d. final magnetization state after application of STT current

**Model:**

We performed micromagnetic simulation using mumax3 [21] where we divided our 50nm x 50nm x 1.5 nm MTJ free layer into 32 x 32 x 1 cells. The cell size is well within the ferromagnetic exchange length $\sqrt{2A_{ex}/\mu_0 M_s^2} \sim 6$ nm. We simulate an inhomogeneous nanomagnet with 10 nm average lateral dimension grains (regions with different anisotropies created by the Voronoi tessellation) and a Gaussian anisotropy variation of 5% standard deviation within grains. Magnetization dynamics was simulated by solving the Landu-Lifshitz-Gilbert-Slonczewski equation in MuMax3 [21]:

$$(1+\alpha^2)\frac{d\vec{m}}{dt} = -\gamma \vec{m} \times \vec{B}_{eff} - \alpha\gamma\left(\vec{m} \times (\vec{m} \times \vec{B}_{eff})\right) - \beta\gamma(\varepsilon - \alpha\varepsilon')(\vec{m} \times (\vec{m}_P \times \vec{m})) \\ + \beta\gamma(\varepsilon' - \alpha\varepsilon)(\vec{m} \times \vec{m}_P) \quad (1)$$

$$\beta = \frac{\hbar J}{edM_s}, \quad \varepsilon = \frac{P\Lambda^2}{(\Lambda^2 + 1) + (\Lambda^2 - 1)(\vec{m} \cdot \vec{m}_p)} \quad (2)$$

Here $\gamma$ is the gyromagnetic ratio, $\alpha$ is the damping, $J$ is the current density along z direction, $d$ is the free layer thickness, $e$ is the electron charge, $\vec{m}$ is the reduced magnetization, $M_s$ is the saturation magnetization, $\vec{m}_p$ is the fixed layer magnetization. We assume Slonczewski parameter, $\Lambda = 1$, secondary spin-torque parameter $\varepsilon' = 0$ and spin polarization $P = 0.5669$.

$\vec{B}_{eff}$ can be expressed as the contribution from several fields:

$$\vec{B}_{eff} = \vec{B}_{anis} + \vec{B}_{demag} + \vec{B}_{stress} + \vec{B}_{exch} + \vec{B}_{thermal} \quad (3)$$

We assume first order uniaxial anisotropy field, $\vec{B}_{anis}$ :

$$\vec{B}_{anis} = \frac{2K_1}{M_s}(\vec{v}.\vec{m})\vec{v} \quad (4)$$

$K_1$ is the first order anisotropy constant and $\vec{v}$ represents the uniaxial anisotropy direction (i.e. perpendicular to plane).

Furthermore, uniaxial stress is applied in plane (perpendicular to the out of plane anisotropy direction) and the stress induced field due to the inverse magnetostriction (Villari) effect, $\vec{B}_{stress}$ can be expressed as:

$$\vec{B}_{stress} = \frac{2K_t}{M_s}(\vec{s}.\vec{m})\vec{s} \quad (5)$$

The effective stress anisotropy constant can be represented by $K_t = \frac{3}{2}\lambda_s\sigma$, where $\lambda_s$ is the saturation magnetostriction and $\sigma$ is the stress amplitude produced by SAW and $\vec{s}$ represents the stress direction. We note that stress induced in the in-plane direction perpendicular to the $\vec{s}$ direction (which is opposite in sign) can add to the $\vec{B}_{stress}$. Since, we do not include this term; our stress estimation is conservative though the qualitative dynamics remains unchanged.

Thermal fluctuation generates $\vec{B}_{thermal}$ in the following manner:

$$\vec{B}_{thermal} = \vec{\eta}\sqrt{\frac{2\mu_0 \alpha k_B T}{M_s \gamma \Omega \Delta}} \tag{6}$$

$\vec{\eta}$ is a randomly generated normal Gaussian distributed vector, $k_B$ is Boltzman constant, $\Omega$ is the cell volume, $\Delta$ is the step size and $\mu_0$ is the magnetic permeability in free space. The simulation parameters adopted are listed in Table I.

Table I: FeGa material properties [22, 23]

| Parameters | $Fe_{81}Ga_{19}$ |
|---|---|
| Saturation magnetostriction ($\lambda_s$) | 350 ppm |
| Gilbert damping ($\alpha$) | 0.015 |
| Saturation magnetization ($M_s$) | $0.8 \times 10^6$ A/m |
| Gyromagnetic ratio ($\gamma$) | $1.76 \times 10^{11}$ rad/Ts |
| Exchange stiffness ($A_{ex}$) | 18 pJ/m |

**Results and Discussion:**

We first simulate a perpendicular homogeneous nanomagnet with an energy barrier ~70 KT with uniform initial magnetization tilted (~2°) from the perpendicular z-axis. This assumption is conservative as thermal noise was found to tilt the mean equilibrium magnetization by ~20°. To investigate the magnetization precession behavior with time varying strain, we excite the nanomagnet with SAW of different frequencies at T= 0 K. In response, the magnetization starts to precess in a cone around the perpendicular axis as seen from Fig. 2a. The precession slowly settles to a mean deflection of ~35° from the perpendicular direction when 100 MPa SAW excitation is applied for a sufficiently long time. However, we note that even a static 200 MPa stress cannot induce any reasonable equilibrium deflection (< 1°), proving that the large deflection from the perpendicular direction is specifically a resonance effect.

The average precession cone deflection angle (polar angle, $\theta$ see Fig. 1) is plotted for different excitation frequency in Fig. 2b for varying stress amplitude. As we can see from the Fig. 2b, the highest deflection angle (resonant point) shifts towards the low frequency with increasing stress amplitude. As we increase the stress amplitude, the magnetization deflects more from the perpendicular direction and the effective field in the perpendicular z-direction decreases. This low effective field in z-direction causes slower magnetization precession at resonance.

Introducing inhomogeneity in nanomagnets (i.e. anisotropy variation between grain) could modify the overall (mean effective) anisotropic field strength along the z-direction and consequently alter the resonance frequency compared to the homogeneous nanomagnets. As different grains have different resonance frequencies, one may expect that the deflection angle vs. frequency is less steep (resonance is not sharp). However, it is likely that the strong exchange interaction forces the individual regions' magnetization to precess nearly in phase. Therefore, the deflection angle vs. frequency (and resonance characteristics) was found to be similar to the homogeneous structure with a mere frequency shift because of mean anisotropy change. Similarly, in the presence of room temperature thermal noise (at T=300 K) the equilibrium magnetization fluctuates randomly producing a higher mean deflection angle. In such a situation, the effective field in the z-direction is lower compared to the T=0 K case. Therefore, the resonant point was found to shift towards low SAW excitation frequency (not shown in Fig. 2b).

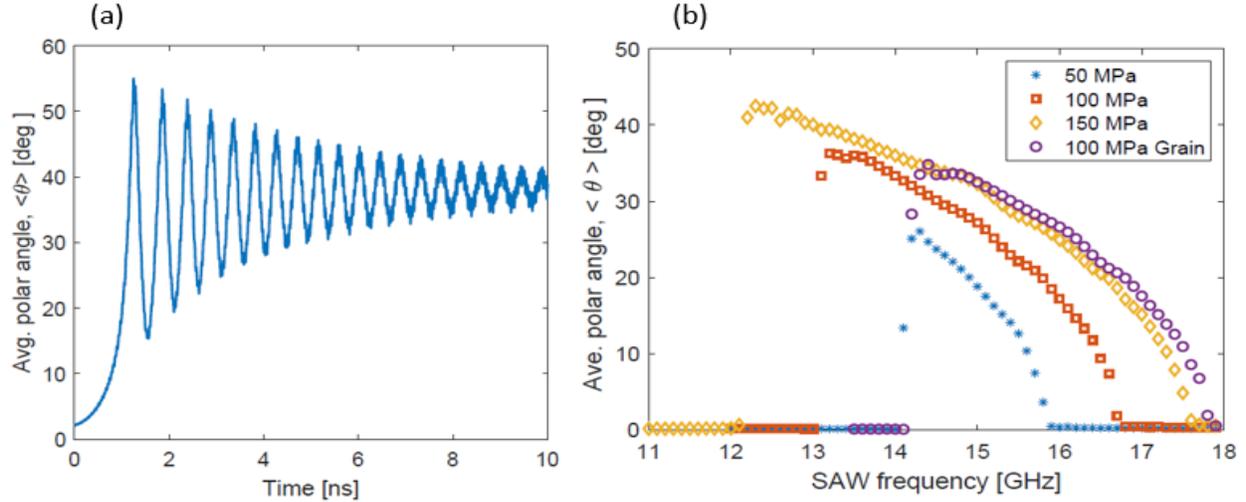

FIG. 2. a. Evolution of magnetization deflection (θ) from the perpendicular z direction for 100 MPa 13.2 GHz SAW b. Average polar angle deflection (θ) for different excitation frequency for varying stress amplitude. Resonance point (highest deflection) shifts towards lower frequency with increasing stress.

The micromagnetic configurations of SAW induced magnetization dynamics are presented in Fig. 3. For the no grain case at T=0 K (Fig. 3a), the spins rotate coherently and eventually settle to an equilibrium cone. Similar behavior is observed for the case with inhomogeneous grains except the spin dynamics is incoherent as spins in different regions precess with different polar angle ($\theta$) with respect to the z-axis (Fig. 3b, 0.99 ns and 1.61 ns). However, at room temperature both homogeneous (Fig. 3c) and inhomogeneous case (Fig. 3d) become incoherent. Notably, for the incoherent cases, the average magnetization deflection can still be high. We next investigate the incoherency and how it affects the magnetization dynamics in SAW driven FMR. Local variation in anisotropy due to material inhomogeneity and thermal perturbation introduces significant incoherency in the nanomagnet as evident from different deflection angles of magnetization precession for different regions with SAW excitation. Notwithstanding the incoherency due to the different deflection angles (polar angle, $\theta$), when driven by resonant SAW that produces a sufficiently large $\theta$, magnetization in the different regions precess almost in phase (in-plane azimuth angle, $\emptyset$). This phase matching of the precessions can produce high net deflection from the perpendicular axis.

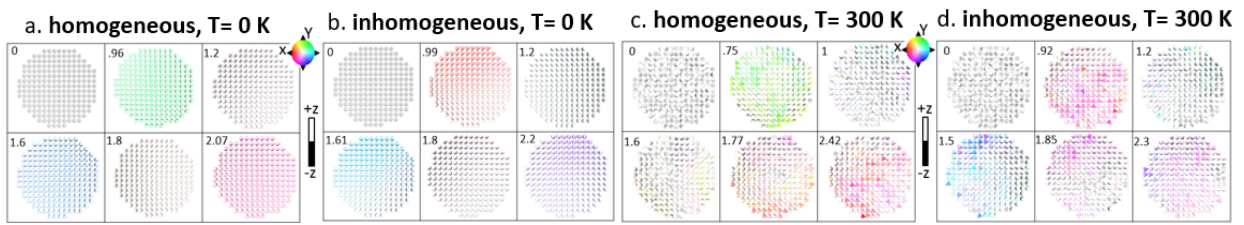

FIG. 3. Evolution of magnetization states with time (shown at the left corner in nanosecond) under the excitation of SAW in a. homogeneous and b. inhomogeneous nanomagnets at T=0 K and c. homogeneous and d. inhomogeneous nanomagnets at T=300 K.

Fig. 4a and 4b plots the spin configuration for inhomogeneous nanomagnet at T=300K, which show the evolution of polar angle ($\theta$) and the in-plane phase angle/azimuth angle ($\emptyset$) for individual spins respectively. Mean polar angle deflection (<$\theta$>) of the magnetization at every time reference is also presented. From Fig 4 b it is evident that spins in different regions are repeatedly precessing almost in phase ($\emptyset$) while producing high mean polar angle (<$\theta$>) (snapshot at .61 ns, 1.35 ns, 2.12 ns in Fig. 4b) but out of phase for low mean polar angle (snapshot at .25 ns, 0.94 ns, 1.71 ns in Fig. 4b). This suggests that high polar deflection between the regions at resonance makes our SAW assisted scheme robust to inhomogeneities.

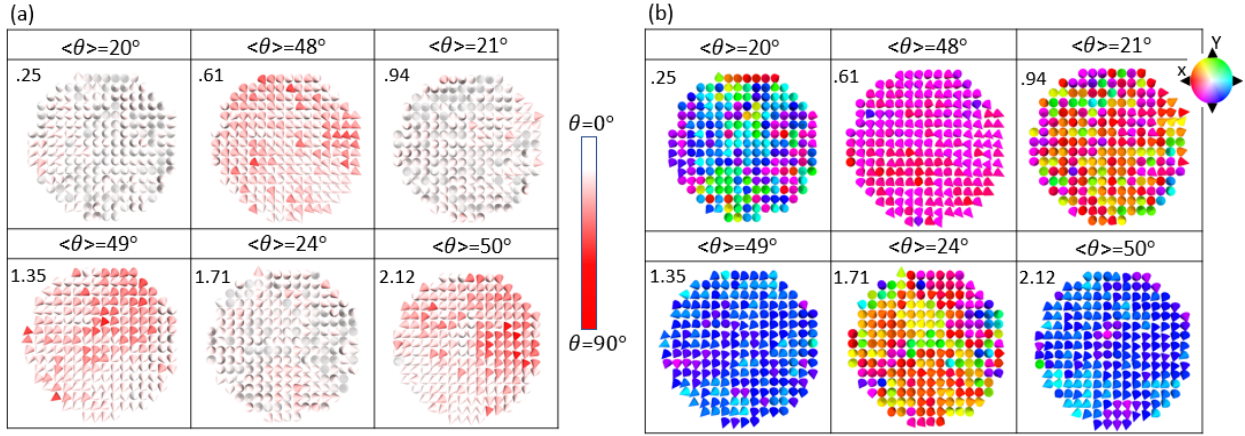

FIG. 4. Spin configuration of inhomogeneous nanomagnet at T=300 K where time in nanoseconds is shown in left corner and average magnetization polar angle (<$\theta$>) is shown for every configuration a. polar angle ($\theta$) magnitude variation in different region spins implies incoherency for both high and low average magnetization polar angle (<$\theta$>) b. in-plane azimuth angle ($\emptyset$) for individual spins shows that the spins are almost in phase while precessing in high polar angle (<$\theta$>) but out of phase while precess in low polar angle .

This large mean deflection in the presence of thermal noise and inhomogeneity improves the efficacy of SAW assisted STT devices. The STT effective torque can be expressed as $\vec{B}_{STT} = \frac{\beta\varepsilon(\vec{m}_P \times \vec{m})}{\gamma}$ , which shows that the torque magnitude is a function of $sin\theta$, where $\theta$ is the angle between fixed layer and free layer magnetization. Thus, the SAW induced high magnetization precession of the free layer can assist in building large STT torque compared to the no SAW case. We investigate the performance of SAW assisted STT switching scheme in the presence of room temperature thermal noise. Here, we assume the SAW simultaneously excites arrays of several MTJs and the STT current writes bits (Fig. 1a) and thus we do not consider a precise synchronization between the SAW and STT application. This is simulated as follows: we excite the nanomagnets with SAW from t=0 to t=5 ns while STT current is applied for 1ns from t=3ns to t=4ns. Therefore, after the withdrawal of STT current, the SAW is still applied for 1ns (t=4 ns to t=5 ns) as the MTJs can be exposed to SAW even after the STT current pulse is withdrawn. We analyze the final magnetization states of the MTJs after 2 ns of SAW withdrawal (t = 7 ns). We assume the threshold for switching to be ~130°, which is conservative. Several switching trajectories for both the case with no grain (no inhomogeneity) and grain (inhomogeneity) are presented in Fig. 5, where the granular nanostructure is sketched in the inset of Fig. 5b. From Fig. 5a and 5b, we can see that the precession cone (polar angle, $\theta$) oscillates between highest and lowest peaks at ~ 1 GHz, which motivates the STT application window of 1 ns, so it coincides with at least one precession peak.

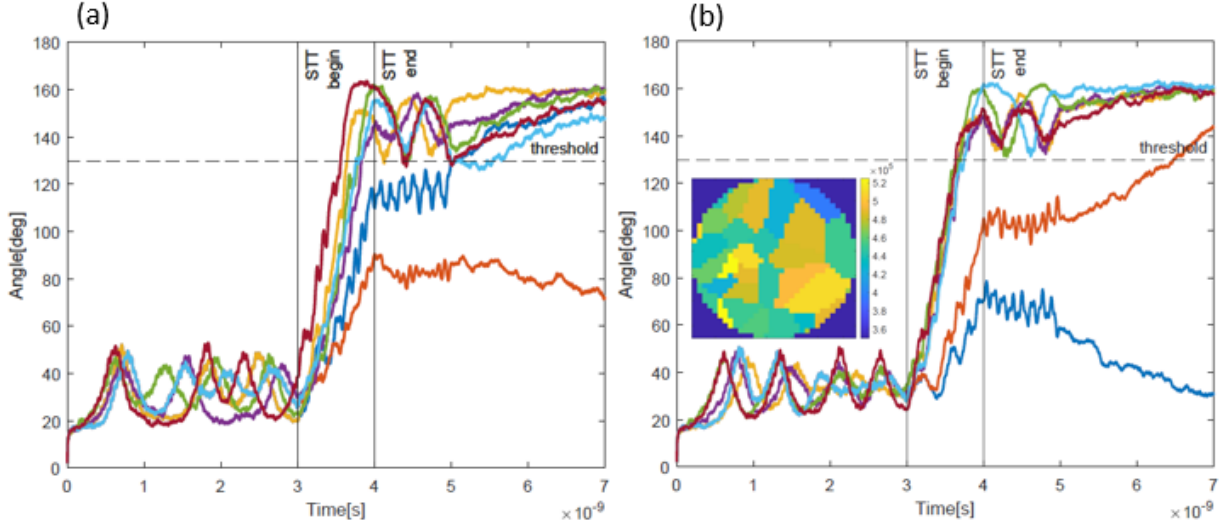

FIG. 5. Switching trajectories of SAW assisted STT a. without grains b. with grains. The inset shows the grain configuration of the nanomagnet and the color bar represents the values of first order anisotropic constant, $K_1$.

Finally, to see the effectiveness of our SAW assisted STT switching scheme we simulate 100 switching trajectories for different stress amplitudes and varying STT current for both homogeneous and inhomogeneous nanomagnets (Fig. 6). For the no SAW case, STT current is applied for 1ns and final magnetization state was evaluated 2 ns after the STT is withdrawn with timing details and reasons for their choice described earlier. From the switching probability curve, it is seen that SAW assisted STT scheme requires less STT current than the no SAW case. While the no SAW case requires a current density of $2.0 \times 10^{11} \ A/m^2$ for switching with 100% probability, in the presence of SAW the STT current can be reduced to $1.4 \times 10^{11} \ A/m^2$. The energy dissipation

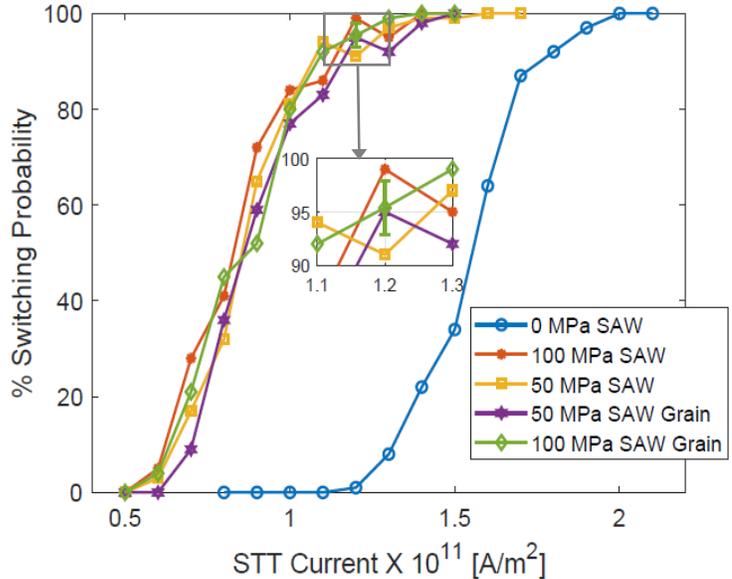

FIG. 6. Switching Statistics with different STT current. Error bar shown in the inset corresponds to one of the data points (100 MPa SAW grain case at $1.2 \times 10^{11}$ A/m$^2$ current) for 1000 simulations

due to SAW excitation is very low. Moreover, energy pumped from one SAW source is amortized over several MTJs. Therefore, energy dissipation is dominated by STT current and our scheme thus provides at least two-times improvement in energy efficiency. The interesting result to be noted here is that regardless of homogeneous or inhomogeneous nanomagnet studied the switching current is decreased by approximately the same amount for the same SAW amplitude. Therefore, the incoherent magnetization precession in granular nanomagnets does not degrade the performance of the SAW assisted STT switching. Moreover, for stress induced change in anisotropy is 4 times less than the total anisotropy barrier for 50 MPa SAW.

## Conclusions:

In summary, we have studied magnetization precession dynamics with time varying stress generated by SAW for nanomagnets with inhomogeneity due to lateral variations in anisotropy as well as thermal noise. When SAW induces FMR in such inhomogeneous nanomagnets, different regions' spins precess nearly in phase at high net magnetization deflection, lowering STT current requited to switch the magnetization. While out of phase precession occurs at low deflection from perpendicular anisotropy axis, the precession in different regions synchronize (are in phase) for high amplitude magnetization deflection from anisotropy axis. Thus, the efficacy of the SAW that produces large resonant deflections does not degrade due to such incoherency in the presence of lateral anisotropy variations and thermal noise.

Technologically, such SAW induced FMR assisted STT memory devices have potential to scale below ~20 nm lateral dimensions even though the concomitantly high anisotropy energy density needed at such small volumes cannot be overcome by static stress. In addition, using well optimized SAW excitation frequency that can maximize the torque on the magnetization when STT is applied and choosing magnetostrictive materials with extremely low damping material one can potentially achieve over an order of magnitude energy reduction in write energy while being able to scale aggressively to very low lateral dimensions. Finally, such SAW induced FMR could have application in low power electronics beyond memory devices [24].

## Acknowledgement:

The work is supported by small grant National Science Foundation SHF Small Grant No. #CCF-1815033 and Virginia Microelectronics Consortium (VMEC) seed grant.

## References:

1. J. C. Slonczewski, Current-driven excitation of magnetic multilayers, J. Magn. Magn. Mater 159, L1 (1996).
2. H. Kubota, A. Fukushima, K. Yakushiji, T. Nagahama, S. Yuasa, K. Ando, H. Maehara, Y. Nagamine, K. Tsunekawa, D. D. Djayaprawira, N. Watanabe, and Y. Suzuki, Quantitative measurement of voltage dependence of spin-transfer torque in MgO-based magnetic tunnel junctions, Nat. Phys. 4, 37 (2008).
3. J. J. Nowak, R. P. Robertazzi, J. Z. Sun, G. Hu, J.-H. Park, J. Lee, A. J. Annunziata, G. P. Lauer, R. Kothandaraman, E. J. O'Sullivan, P. L. Trouilloud, Y. Kim, and D. C. Worledge, Dependence of voltage and size on write error rates in spin-transfer torque magnetic random-access memory, IEEE Magn. Lett. 7, 3102604 (2016).
4. Z. Zhao, M. Jamali, N. D'Souza, D. Zhang, S. Bandyopadhyay, J. Atulasimha, and J.-P. Wang, Giant voltage manipulation of MgO-based magnetic tunnel junctions via localized anisotropic strain: A potential pathway to ultra-energy-efficient memory technology, Appl. Phys. Lett. 109, 092403 (2016).
5. P. Li, A. Chen, D. Li, Y. Zhao, S. Zhang, L. Yang, Y. Liu, M. Zhu, H. Zhang, and X. Han, Electric field manipulation of magnetization rotation and tunneling magnetoresistance of magnetic tunnel junctions at room temperature, Adv. Mater. 26, 4320 (2014).

6. Y. Shiota, S. Miwa, T. Nozaki, F. Bonell, N. Mizuochi, T. Shinjo, H. Kubota, S. Yuasa, and Y. Suzuki, Pulse voltage-induced dynamic magnetization switching in magnetic tunneling junctions with high resistance-area product, Appl. Phys. Lett. 101, 102406 (2012).
7. D. Bhattacharya and J. Atulasimha, Skyrmion-mediated voltage-controlled switching of ferromagnets for reliable and energy-efficient two-terminal memory, ACS Appl. Mater. Interfaces 10, 17455 (2018).
8. D. Labanowski, A. Jung, and S. Salahuddin, Power absorption in acoustically driven ferromagnetic resonance, Appl. Phys. Lett. 108, 022905 (2016).
9. M. Weiler, L. Dreher, C. Heeg, H. Huebl, R. Gross, M. S. Brandt, and S. T. B. Goennenwein, Elastically driven ferromagnetic resonance in nickel thin films, Phys. Rev. Lett. 106, 117601 (2011).
10. V. Sampath, N. D'Souza, D. Bhattacharya, G. M. Atkinson, S. Bandyopadhyay, and J. Atulasimha, Acoustic-wave-induced magnetization switching of magnetostrictive nanomagnets from single-domain to nonvolatile vortex states, Nano Lett. 16, 5681 (2016).
11. L. Thevenard, I. S. Camara, S. Majrab, M. Bernard, P. Rovillain, A. Lemaître, C. Gourdon, and J.-Y. Duquesne, Precessional magnetization switching by a surface acoustic wave, Phys. Rev. B 93, 134430 (2016).
12. I.S. Camara, J.-Y. Duquesne, A. Lemaître, C. Gourdon, and L. Thevenard, Field-free magnetization switching by an acoustic wave, Phys. Rev. Appl. 11, 014045 (2019).
13. S. Mondal, M. A. Abeed, K. Dutta, A. De, S. Sahoo, A. Barman, and S. Bandyopadhyay, Hybrid magnetodynamical modes in a single magnetostrictive nanomagnet on a piezoelectric substrate arising from magnetoelastic modulation of precessional dynamics, ACS Appl. Mater. Interfaces 10, 43970 (2018).
14. Y. Yahagi, B. Harteneck, S. Cabrini, and H. Schmidt, Controlling nanomagnet magnetization dynamics via magnetoelastic coupling, Phys. Rev. B 90, 140405(R) (2014).
15. V. S. Vlasov, A. M. Lomonosov, A. V. Golov, L. N. Kotov, V. Besse, A. Alekhin, D. A. Kuzmin, I. V. Bychkov, and V. V. Temnov, Magnetization switching in bistable nanomagnets by picosecond pulses of surface acoustic waves, Phys. Rev. B 101, 024425 (2020).
16. A. Roe, D. Bhattacharya, and J. Atulasimha, Resonant acoustic wave assisted spin-transfer-torque switching of nanomagnets, Appl. Phys. Lett. 115, 112405 (2019).
17. T. Thomson, G. Hu, and B. D. Terris, Intrinsic distribution of magnetic anisotropy in thin films probed by patterned nanostructures, Phys. Rev. Lett. 96, 257204 (2006).
18. J. M. Shaw, S. E. Russek, T. Thomson, M. J. Donahue, B. D. Terris, O. Hellwig, E. Dobisz, and M. L. Schneider, Reversal mechanisms in perpendicularly magnetized nanostructures, Phys.Rev. B 78, 024414 (2008).
19. J. M. Shaw, H. T. Nembach, and T. J. Silva, Roughness induced magnetic inhomogeneity in Co/Ni multilayers: Ferromagnetic resonance and switching properties in nanostructures, J. Appl. Phys. 108, 093922 (2010).
20. D. Winters, M. A. Abeed, S. Sahoo, A. Barman, and S. Bandyopadhyay, Reliability of magnetoelastic switching of nonideal nanomagnets with defects: a case study for the viability of straintronic logic and memory, Phys. Rev. Appl. 12, 034010 (2019).
21. A. Vansteenkiste, J. Leliaert, M. Dvornik, M. Helsen, F. G. -Sanchez, and B. V. Waeyenberge, The design and verification of MuMax3, AIP Advances 4, 107133 (2014).
22. D. B. Gopman, V. Sampath, H. Ahmad, S. Bandyopadhyay, and J. Atulasimha, Static and dynamic magnetic properties of sputtered Fe–Ga thin films, IEEE Tran. Magn. 53, 6101304 (2017).
23. A. E. Clark, M. Wun-Fogle, J. B. Restorff, T. A. Lograsso, Magnetostrictive properties of galfenol alloys under compressive stress, Mater. Trans. 43, 881 (2002).

24. A. C. Chavez, J. D. Schneider, A. Barra, S. Tiwari, R. N. Candler, and G. P. Carman, Voltage-controlled ferromagnetic resonance of dipole-coupled $Co_{40}Fe_{40}B_{20}$ nanoellipses, Phys. Rev. Appl. 12, 044071 (2019).